# A minor planet in an outer resonance with Uranus

Daniel Bamberger
Northolt Branch Observatories
Marburg, Germany
https://orcid.org/0000-0002-9138-2942

K Ly
University of California, Los Angeles
Los Angeles, CA, USA
https://orcid.org/0009-0004-9268-9796

Sam Deen
Deep Random Survey
Simi Valley, CA, USA
https://orcid.org/0009-0004-6814-5449

Elvis Oliveira Mendes
Instituto de Física, Universidade Federal Fluminense
Niterói, Rio de Janeiro, Brazil

Abstract

We have located archival observations of the centaur 2015 $OU_{194}$ from 2017 and 2018, which extend its data-arc length from 1.0 to 3.5 years. We show that it is in an outer 3:4 mean motion resonance with Uranus, henceforth referred to as $U_{3/4}$. The resonance is stable from at least 1000 kyrs in the past till 500 kyrs in the future.

We find no mention in the literature of known objects in this resonance, or in any other resonance between the orbits of Uranus and Neptune. Looking for additional candidates, we find that 2013 $RG_{98}$ also stays in $U_{3/4}$ for several hundred kyrs around the present epoch. A third candidate, 2014 $NX_{65}$, is strongly influenced by Neptune.

## 1. Introduction

Minor planets between Saturn and Neptune, known as centaurs, are characterized by the chaotic and unstable nature of their orbits.

A mean motion resonance (MMR) occurs when the quantity $pn-qn'$ is approximately zero, where $p$ and $q$ are positive integers, and $n$ and $n'$ are the mean motion frequencies of the small body and the planet, respectively (Li et al. 2018).

The longitude of pericenter is given by $\varpi = \omega + \Omega$, and the mean longitude of the minor planet is defined as $\lambda = M + \omega + \Omega$, where $\omega$, $\Omega$, and $M$ are the argument of pericenter, longitude of the ascending node, and mean anomaly, respectively. A small body is said to be in a q:p mean motion resonance when the resonant angle $\phi = p\lambda - q\lambda' - (p-q)\varpi$ librates (Morais & Namouni 2017).

All known MMRs with Uranus are located between the orbits of Uranus and Saturn ($q \geq p$). Examples that have been discussed in the literature are 2000 QC$_{243}$ in a 5:4 MMR, and 2001 XZ$_{255}$ in a 4:3 MMR (Masaki & Kinoshita 2003).

## 2. Observations and orbital elements

The discovery of 2015 OU$_{194}$ was announced on 14 May 2025,[1] with additional observations being published on 27 Jun. 2025.[2] It was first observed at Subaru Telescope, Maunakea, on 12 Jul. 2015, and at the time of writing, the last observation was from exactly one year later, 12 Jul. 2016. A search of the Subaru Telescope archive (Gywn et al. 2012) revealed that it has been observed again on 16 Sept. 2017 and 7 Dec. 2018.[3] The observations have been submitted to the Minor Planet Center.

From a total of 34 observations, the following nominal orbit, with a mean residual of 0.15", is computed using Bill Gray's Find_Orb software.[4] The epoch is JD 2458459.5 (2018 Dec. 7.0 TT):

a = 23.32719 ± 0.00189 au, e = 0.089211 ± 0.000105, Incl. = 11.983030 ± 0.000012°, Peri. = 38.194 ± 0.016°, Node = 243.8055 ± 0.0007°, M = 54.230 ± 0.010°, q = 21.246153 ± 0.000898 au, Q = 25.40824 ± 0.00448 au, P = 112.67 years.

The observations contain 31 multiband photometry measurements. The calculated average colors of these are g-r = 0.79, r-i = 0.70, i-z = 0.24, z-Y = 0.44 (in Subaru Telescope filters; no uncertainties reported), colors which are roughly average for an inactive centaur or trans-Neptunian object (Bolin 2021; Schwamb 2019).

## 3. Simulations and results

With the extended data-arc, the stability of the resonance can be characterized. Using the SOLEX/EXORB package (Vitagliano 1996),[5] we create a pair of clones at 3-sigma on either side of the mean, and compare their evolution between -1000 kyrs and +500 kyrs. All three remain trapped in U$_{3/4}$, with the resonance angle $\phi = 4\lambda - 3\lambda' - \varpi$ librating at 180° and with an amplitude that rarely exceeds 200°. The low amplitude indicates that the resonance may be quite stable. Before -1000 kyrs, the behavior of the clones diverges, and while the nominal orbit remains in resonance at -1100 kyrs, the two clones do not. A more sophisticated analysis, necessary to determine its dynamical lifetime, is beyond the scope of this paper.

## 4. Additional candidates for a $U_{3/4}$ resonance

To find out whether 2015 $OU_{194}$ is unique among the known minor planets, we also ran simulations for the nominal orbit of 2013 $RG_{98}$ (a=23.35 au, e=0.17, i=46.0°, data-arc 8.2 years) and 2014 $NX_{65}$ (a=23.02 au, e=0.20, i=11.4°, data-arc 10.1 years), which are the only other known objects with low eccentricity (e<0.3) and semi-major axes within 0.3 au of $U_{3/4}$ at 23.28 au.[6]

We find that the nominal orbit of 2013 $RG_{98}$ also librates in $U_{3/4}$ from around -400 kyrs until +470 kyrs, but at a much larger amplitude that reaches nearly 360°. The resonance has been intermittent before -400 kyrs, and it will break at +470 kyrs.

2014 $NX_{65}$ is not resonant, but it interacts with both Uranus and Neptune. Its nominal orbit switches between $U_{3/4}$, no libration at all, and an inner 3:2 resonance with Neptune, with each resonant phase lasting for tens of kyrs.

---

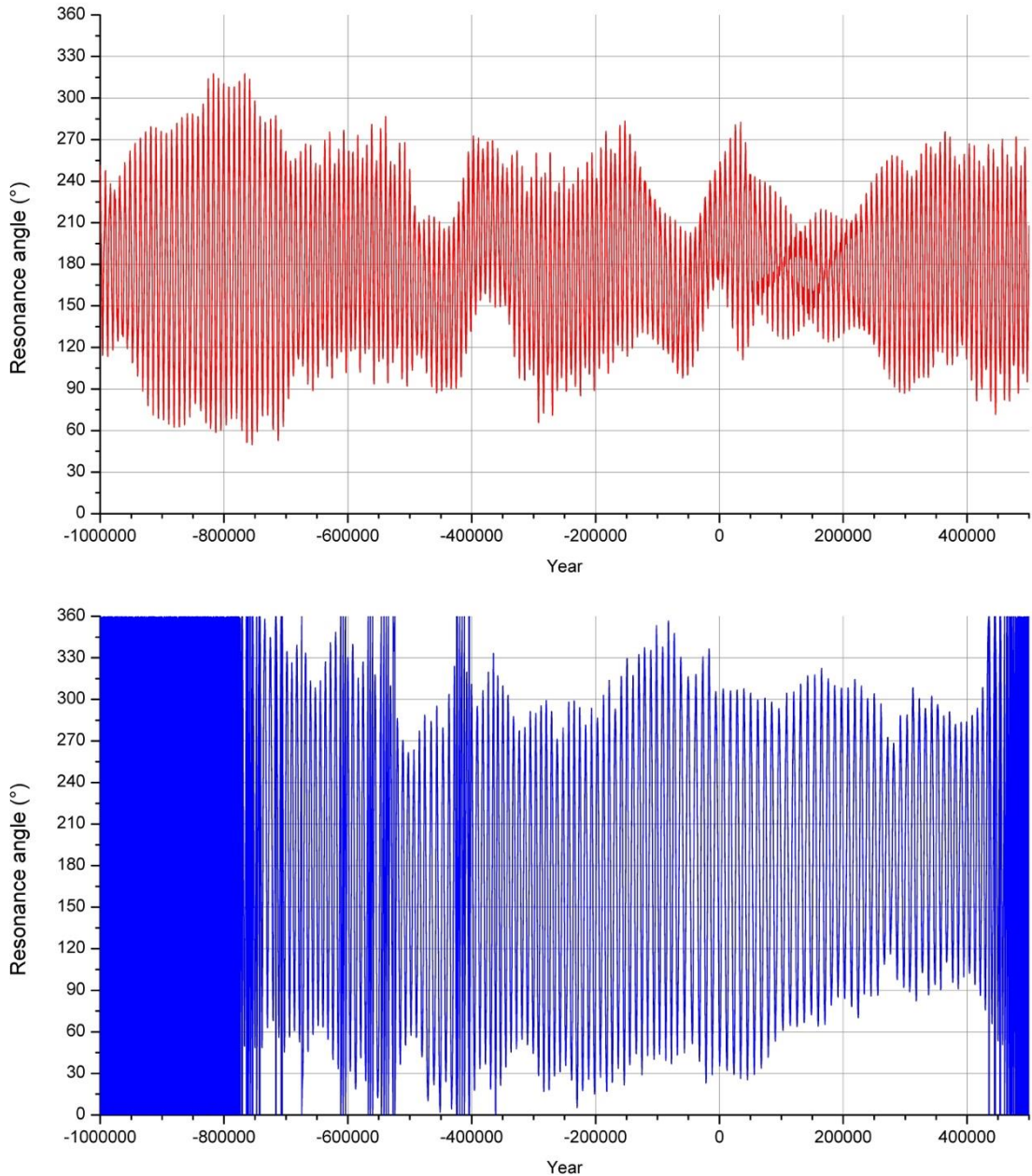

Figure 1: *The resonance angle for the nominal orbits of 2015 OU$_{194}$ (top, red) and 2013 RG$_{98}$ (bottom, blue), between -1000 kyrs and +500 kyrs from the present day. The dataset used to generate this figure is available for download in machine-readable format from the online journal.*

## 5. Conclusions

Among the known objects located close to an outer resonance with Uranus, 2015 OU$_{194}$ is unique for its nearly circular orbit, low libration angle, and the apparent stability of a type of resonance that, as far as we know, has not been seen before.